\documentclass[aps,prl,twocolumn]{revtex4-1}

\usepackage{hyperref}
\usepackage{color}
\usepackage{graphicx,psfrag, subfigure}
\usepackage{bbold}

\usepackage{mathtools}
\usepackage{braket}
\newcommand{\bigslant}[2]{{\raisebox{.2em}{$#1$}\left/\raisebox{-.2em}{$#2$}\right.}}

\usepackage{array}

\usepackage{color}
\usepackage{eufrak}
\usepackage{graphicx,psfrag, subfigure}
\usepackage{amsmath}
\usepackage{hyperref}
\DeclareMathAlphabet{\mathpzc}{OT1}{pzc}{m}{it}
\usepackage{blkarray}
\usepackage[applemac]{inputenc}

\usepackage{yfonts}
\usepackage{mathrsfs}
 \usepackage{relsize}
\usepackage{tabularx}

\def\M{M_{{\rm{pl}}}}

\def\be{\begin{equation}}
\def\ee{\end{equation}}
\def\bea{\begin{eqnarray}}
\def\eea{\end{eqnarray}}

\def\be{\begin{equation}}
\def\ee{\end{equation}}
\def\bea{\begin{eqnarray}}
\def\eea{\end{eqnarray}}

\usepackage{dsfont}
\usepackage[mathscr]{eucal}

\begin{document}

\title{Axionic Band Structure of the Cosmological Constant}

\author{Thomas C.~Bachlechner} \affiliation{Physics Department and Institute for Strings, Cosmology and Astroparticle Physics, Columbia University, New York, NY 10027, USA}
\vskip 4pt

\begin{abstract}
We argue that theories with multiple axions generically contain a large number of vacua that can account for the smallness of the cosmological constant. In a theory with $N$ axions, the dominant instantons with charges $\mathscr Q$ determine the discrete  symmetry of vacua. Subleading instantons  break the leading periodicity and lift the vacuum degeneracy. For generic integer charges the number of distinct vacua is given by $\sqrt{\det({\mathscr Q^\top \mathscr Q})}\propto e^N$. Our construction motivates the existence of a landscape with a vast number of vacua in a large class of four-dimensional effective theories.

\end{abstract}

\maketitle

\section{Introduction}
Einstein's field equations famously couple the vacuum energy density $\rho_{\Lambda}$ to the curvature of spacetime. Supernova data and cosmic microwave background observations are well described by general relativity with a small, positive vacuum energy $\rho_{\Lambda}\approx 10^{-123}\M^4$ \cite{Perlmutter:1998np,Riess:1998cb,Ade:2015xua}. However, known contributions to the vacuum energy 
correct any bare cosmological constant by at least $10^{-64}\M^4$ \cite{Weinberg:1988cp,Carroll:2000fy}. This vast discrepancy in scales is known as the cosmological constant problem. One of the best motivated approaches to this apparent fine-tuning is to study a large number of populated vacua and consider selection effects: any experiment we perform is subject to a selection bias that skews the distribution of observed outcomes \cite{Banks:1984tw,Linde:1984ir,Barrow:1988yia,Weinberg:1987dv,Polchinski:2006gy,Bousso:2006nx}. For example, Weinberg first pointed out that galaxies will not form unless $|\rho_{\Lambda}|\lesssim 10^{-121}\M^4$.  In an eternally inflating universe with a diverse vacuum structure, the selection effect would be manifest in the form of a surprisingly small vacuum energy \cite{Bousso:2007kq,DeSimone:2008bq,Bousso:2007er,Clifton:2007bn}.

Brown and Teitelboim (BT) suggested that a single four-form field strength might cancel a bare cosmological constant to arbitrary accuracy \cite{Abbott:1984qf,Brown:1987dd,Brown:1988kg,Feng:2000if}. Independently, Banks, Dine and Seiberg (BDS) noted that theories with a single axion, but irrational decay constants, also lead to a large landscape of vacua \cite{Banks:1991mb}. Unfortunately, both proposals suffer from cosmological problems and resist an embedding in quantum gravity \cite{Abbott:1984qf,Brown:1987dd,Brown:1988kg,Banks:1991mb,Feng:2000if,Banks:2000pj,Douglas:2006es,Bousso:2007gp,Kallosh:2014vja}. 
The BT mechanism was generalized to the multiple field strengths that generically arise in string theory, which led to the discovery of a string landscape that famously allows for as many as $10^{500}$ vacua, accommodates a realistic cosmology and contains the desired selection effect of vacuum energies \cite{Bousso:2000xa}. 
In this work we extend the BDS approach to theories of multiple axions, such as string compactifications with fixed fluxes, and discover a field theory landscape with an exponential number of vacua and consistent cosmological history. Crucially, our construction may be embedded in a theory of quantum gravity and provides a new framework for the study of vacua in the string landscape.

\begin{figure}
  \centering
  \includegraphics[width=.48\textwidth]{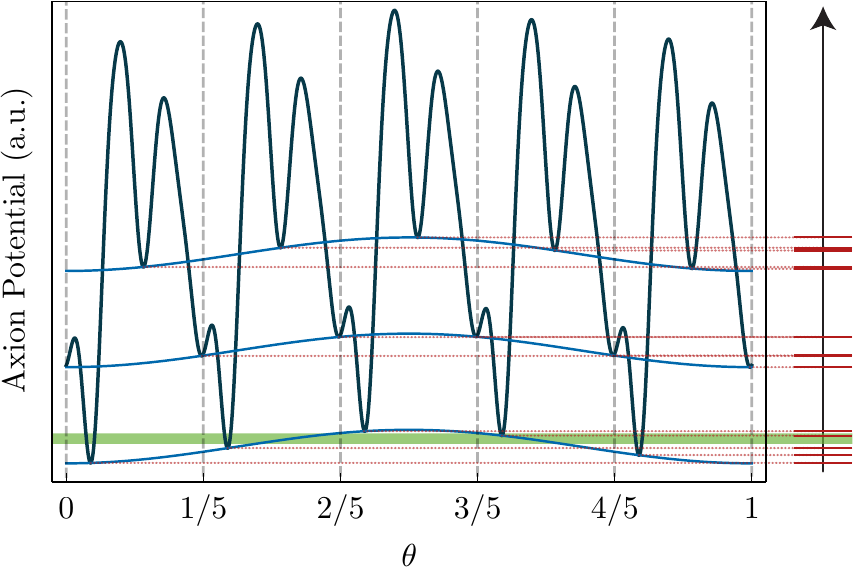}
  \caption{\small Left: Axion potential (dark blue) and a subleading contribution (light blue). Right: Vacuum energies. 
 Selection effects may lead to preferred  vacua in the green band.}\label{fig:bandplot}
\end{figure}

The leading contributions to the axion potential are invariant under some discrete translations $\mathscr Q_{\, i}^a\theta^i\rightarrow  \mathscr Q _{\, i}^a\theta^i+1$. This shift symmetry is broken by subleading contributions to a  periodicity $\theta^i\rightarrow \theta^i+1$. The fundamental region  of the leading potential has a volume of size $1/\sqrt{\det(\mathscr Q^\top \mathscr Q)}$, such that after including the remaining subleading terms, the vacuum energies are split into about $ {\mathcal N}_{\text{r}}\propto \sqrt{\det(\mathscr Q^\top \mathscr Q)}\propto   e^{c N}$ distinct vacua, for some constant $c$. If the charges of the dominant instantons do not form a primitive basis of the integer lattice, we find $c>0$, leading to an exponentially large number of non-degenerate vacua. For example, consider the case of diagonal charges, ${\mathscr Q}^{a}_{\, i}=n \delta^{a}_{\, i}$. The leading potential is invariant under discrete shifts $\theta^i\rightarrow \theta^i+1/n$, but this symmetry is broken by subleading terms to the original unit periodicity, giving rise to $n^N$ distinct vacua. This basic mechanism is illustrated in Figure \ref{fig:bandplot} for the one-axion case.

\section{The vacua of $N$ Axions}
Consider a theory of $N$ axions $\theta^i$, whose continuous shift symmetries are broken only by non-perturbative effects to the discrete shift symmetry $\theta^i\rightarrow \theta^i+1$. The instanton effects  generate an axion potential of the form
\be \label{Lone}
{\mathcal L}={1\over 2}K_{ij} \partial\theta^i\partial\theta^j - \sum_{a} \Lambda_a^4 \left[1-\cos\left(2\pi {\cal Q}^a_{\, i}\theta^i +\delta^a\right) \right]-V_0\,,
\ee
where the sum runs over all contributions to the axion potential, $V_0$ is some constant energy density, $K_{ij}$ is the field space metric and ${\cal Q}^a$ denotes the integer charge vector of the $a^{\text{th}}$ instanton contribution. 

We are interested in the vacuum distribution in this theory. In general, the location of  minima is difficult to obtain: the critical point equation consists of $N$ coupled, non-linear equations. To simplify the problem, let us denote the number of terms that give rise to the leading contributions by $P$. We can now decompose the charge matrix according to that choice:
\be  \label{squarechoice}
{\mathcal Q}^a\big|_{a=1,\dots}= \left(\begin{tabular}{c} $\mathscr Q$\\ $ Q_\text{\text{r}}$\end{tabular}\right)^a\bigg|_{a=1,\dots}\, ,
\ee
where $\mathscr Q$ is a full rank, rectangular $P\times N$ matrix and $Q_\text{\text{r}}$ is a rectangular matrix of rank $R$, containing all remaining charges. 
The leading potential is given by
\bea
V_{\mathscr Q}=\sum_{a=1}^P \Lambda_a^4 \left[1-\cos\left(2\pi {\mathscr Q}^a\boldsymbol \theta +\delta^a\right) \right]\,,
\eea
and we call the typical scale of this potential $\Lambda_{\mathscr Q}^4$.
The charges ${\mathscr Q}$ specify the periodicity of vacua in terms of the lattice ${\mathscr L}(\mathbf B_{\mathscr Q})$ with basis $\mathbf B_{\mathscr Q}$, that is
\bea\label{lQ}
 {\mathscr L} (\mathbf B_{\mathscr Q})&\equiv &\left\{  \mathbf B_{\mathscr Q}\mathbf n~|\mathbf n\in  \mathbb Z^N\right\}
\nonumber \\&=&\bigcap_{a=1}^P \left\{\boldsymbol \theta|  {\mathscr Q}^a \boldsymbol \theta \in {\mathbb Z}\right\} \,.
\eea
In the special case of a square matrix ${\mathscr Q}$, the periodicity is generated by the basis $ {\mathscr Q}^{-1}$.
To make the periodicity of the potential manifest we now perform the $GL(N)$ transformation,
\be\label{phidef}
\boldsymbol \phi=\mathbf B^{-1}_{\mathscr Q}\boldsymbol \theta\,,
\ee
such that the leading potential is invariant under the discrete shifts  $\phi^i\rightarrow \phi^i+1$. Let us call the number of distinct vacua of the leading potential ${\cal N}_{\mathscr Q}\ge 1$. These vacua are located at field values
\be\label{eq:phivacua}
\boldsymbol \phi^*_{\alpha,\mathbf n} =\boldsymbol \phi^*_\alpha+\mathbf n\,,~~{\mathbf n}\in \mathbb Z^N\,,
\ee
where the index $\alpha$ labels distinct vacua. The scale of the vacuum energies is set by the leading dynamical scales, i.e. $V(\boldsymbol \phi^*_\alpha)\sim \Lambda_{\mathscr Q}^4$.

\section{The Axionic Band Structure}
While the leading terms in the potential exhibit degenerate vacua that are invariant under discrete shifts on the lattice $ {\mathscr L} (\mathbf B_{\mathscr Q})$, this symmetry  is broken by the subleading potential. As in (\ref{lQ}), let us refer to the basis that generates the lattice of rank $R$, under which the subleading potential is invariant, as $\mathbf B_{\text{r}}$. In terms of the field $\boldsymbol \phi$, the subleading potential is symmetric under the discrete shifts \footnote{Here, the inverse only acts on the $R$-dimensional subspace on which the lattice ${\mathscr L}(\mathbf B_{\text{r}})$ is defined.}
\be
\mathbf B_{\text{r}}^{-1}\mathbf B_{\mathscr Q}\boldsymbol \phi\rightarrow \mathbf B_{\text{r}}^{-1}\mathbf B_{\mathscr Q}\boldsymbol \phi + \mathbf n\,,~~\mathbf n \in \mathbb Z^N\,.
\ee
Therefore, when including the subleading terms, the vacua of $V_{\mathscr Q}$ are periodic on the lattice generated by the basis $ \mathbf B_{\text{r}}^{-1}\mathbf B_{\mathscr Q}$, modulo $ \mathbb Z^{R}$, 
\be
{\mathscr L}( {\mathcal B})=\bigslant{\left\{   ({\mathbf  B}^{-1}_{\text{r}}\mathbf B_{\mathscr Q}) {\mathbf n}|~\mathbf n\in   \mathbb Z^N\right\}}{  \mathbb Z^R}\,,
\ee
where we call the corresponding basis ${\mathcal B}$. Each of the ${\cal N}_{\text{r}}$ sites of the lattice ${\mathscr L}({\mathcal B})$, that are located within the unit $R$-hypercube, corresponds to a distinct, non-degenerate vacuum. The fundamental parallelepiped of the lattice was defined through a quotient  by $\mathbb Z^R$, so it is  a tiling of the unit $R$-hypercube. Therefore, the number of distinct vacua is given by the inverse volume of the fundamental parallelepiped,
\be
{\cal N}_{\text{r}}= \sqrt{\det{({\mathcal B}^\top {\mathcal B}})^{-1}}\,.
\ee
This is the main result of our work. Each of the ${\cal N}_{\mathscr Q}$ vacua of the leading potential is split into an energy band of width $\Lambda_{\text{r}}^4$, containing  ${\cal N}_{\text{r}}$ vacua. We refer to this vacuum distribution as the {\it axionic band structure}. If any of the vacua of the leading potential vacua are within about $\Lambda_{\text{r}}^4$ of zero, there exist vacua with energies as low as $\Lambda_{\text{r}}^4/{\cal N}_{\text{r}}$.

Returning to the simple example of $P=N$ leading terms and we take $ Q_{\text{r}}$ to generate the integer lattice, such that $\mathbf B_{\text{r}}=\mathbb 1$, we immediately find the number of distinct vacua to be
\be\label{numbervacua}
{\cal N}_{\text{r}}=\sqrt{\det({\mathscr Q}^\top \mathscr Q)}\,.
\ee

\section{Random Axion Theories}

Naively, it might appear unlikely that the axionic bands contain a large number of non-degenerate vacua. In the one-axion case, a large number can only be achieved by an equally large tuning of two axion decay constants to an almost irrational ratio \cite{Banks:1991mb}. We now consider an ensemble of random multi-axion theories defined through the measure on the space of charge matrices ${\cal Q}$. In particular, we will consider charge matrices with entries consisting of independent, identically distributed (i.i.d.) random integers. This choice is motivated from explicit flux compactifications on Calabi-Yau manifolds \cite{Bachlechner:2014gfa,Long:2014fba} 
or from gravitational instantons \cite{Giddings:1987cg,ArkaniHamed:2007js}. 
Even though the charge matrix may be sparse, it rapidly approaches its universal limit when a small fraction $\gtrsim 3/N$ of  entries are non-vanishing \cite{Bachlechner:2014gfa,wood2012}. In the universal limit, the matrix ${\cal Q}^\top {\cal Q}$ is well described by the  Gaussian orthogonal Wishart ensemble, i.e. the ensemble of matrices $\mathbf W$
\be
\mathbf W=\mathbf A^\top\cdot \mathbf A\, ,
\ee
where the entries of $\mathbf A$ are real, i.i.d. random numbers with variance $\sigma^2$ and vanishing mean. For a sparse, square matrix $\mathbf A$ with $\delta N$ non-vanishing integer entries the variance in the universal regime is given by
\be
\sigma^2\approx {\delta N\over N^2}\gtrsim {3\over N}\,.
\ee
The determinant of Wishart matrices follows a product chi-squared distribution, so that the expected value of the determinant is given by \cite{goodman1963,Bachlechner:2015qja}
\be\label{deteq}
\left\langle\det({\cal Q}^\top {\cal Q})\right\rangle=\sigma_{\cal Q}^{2N}\Gamma(N+1)\,.
\ee
Let us now obtain a simple estimate for the number of non-degenerate vacua. The dynamical scales $\Lambda_a^4$ are generated non-perturbatively, so we expect them to be  distributed uniformly on a logarithmic scale. This large hierarchy implies a small number of leading terms, so we can take $P=N$. In the context of string compactifications,  the scales are set by $\Lambda_{a}^4\sim \exp({-{\cal Q}^a_{\,i}\tau^i})$ for some volumes $\tau^i$. Naively, one might worry that the leading charges are small and form a primitive basis for the integer lattice. However, depending on the volumes and other possible constraints on the  charges, the dominant contributions in general are non-trivial.
On the other hand, there is a vast number of subleading instantons with unconstrained charges. These charges do form a primitive basis for the integer lattice, so that we can take $\mathbf B_{\text{r}}=\mathbb 1$.  With (\ref{numbervacua}) and (\ref{deteq}) we immediately obtain the expected number of non-degenerate vacua
\bea
\left\langle {\cal N}_{\text{r}}^2\right\rangle&=&\sigma_{{\cal Q}}^{2N}\Gamma(N+1)\gtrsim \sqrt{2\pi N}  \left({3\over e}\right)^N \,.
\eea
For generic, square charge matrices $\mathscr Q$ with more than about $ 3 N$ non-vanishing integer entries, there exists an exponentially large number of non-degenerate vacua. To give a concrete example, in the presence of $500$ axions, a charge matrix with $2\%$ non-vanishing integer entries suffices to generate more than $ 10^{120}$ distinct vacua.

\section{Cosmological Considerations}
In the previous sections we argued for the existence of an exponentially large number of vacua in theories containing multiple axions. However, a successful theory for the observed value of the cosmological constant not only realizes a classically stable vacuum in the right energy range, but also connects to a realistic cosmology. First, the theory must admit a consistent quantum gravity completion and accommodate (metastable) vacua at energies vastly smaller than the natural scale of the theory. Second, vacua with small cosmological constant are stable on time-scales of the age of the universe. Finally, the theory allows for sufficient energy to drive inflation and reheating. We now briefly discuss these considerations in turn.

We considered generic theories of multiple axions and demonstrated the existence of a vast number of vacua, distributed uniformly over an energy range given by $\Lambda_{\text{r}}^4$ in a potential with typical scale $\Lambda_{\mathscr Q}^4\gtrsim\Lambda_{\text{r}}^4$. Therefore, classically stable vacua with small vacuum energy are not atypical. Due to the discrete shift symmetry in the axion sector, the vacuum distribution is decoupled from high energy physics. However, a domain wall that interpolates between two vacua may have access to high-energy degrees of freedom that change the effective axion potential. 
The extra degrees of freedom do not change the axion symmetry and therefore the number of vacua in each axionic band is unchanged. Finally, a large canonical field displacement during a vacuum transition may be in conflict with quantum gravity or resist an embedding in string theory \cite{ArkaniHamed:2006dz,Cheung:2014vva,Banks:2003sx}. The potential is approximately periodic under discrete shifts $\phi^i\rightarrow \phi^i+1$. Therefore, the canonical separation between  two vacua cannot exceed $\sqrt{N} \xi_N$, where $\xi_N^2$ is the largest eigenvalue of the kinetic matrix for the axions $\boldsymbol \phi$. While the precise form of quantum gravitational constraints on field ranges are currently under  debate, the displacement $ \sqrt{N} \xi_N$ may well be sub-Planckian and therefore decoupled from quantum gravity \cite{Bachlechner:2015qja,Heidenreich:2015wga}. 
We conclude that theories containing multiple axions can accommodate classically stable solutions with exponentially small vacuum energies.

Let us now ask whether the vacua are sufficiently stable against  decay. Given the age of our universe, a sufficient condition for stability is  a small decay rate, $\Gamma\ll e^{-10^3}\M^4$.
In general, it is difficult to estimate the tunneling rate between distinct vacua of the leading potential $V_{\mathscr Q}$, but we can obtain an upper limit for the  tunneling rate between the vacua split by the subleading potential contributions.
Using the thin-wall approximation and neglecting gravity, the  decay rate for tunneling between two (local) minima located at $\Phi_i$ and $\Phi_f$ is given by $\Gamma\sim e^{-B}$ where, \cite{Coleman:1977py,Coleman:1980aw}
\be
B= {27\pi^2 \sigma^4\over 2\epsilon^3}\,.
\ee
Here, the surface tension of the Coleman-De Luccia instanton is given by
\be
\sigma=\int_{\Phi_i}^{\Phi_f}d\Phi ~\sqrt{2[V(\Phi)-V(\Phi_i)]}\,,
\ee
the integration is performed along the path of extremal action and  $\epsilon$ is the difference in energy density between the two vacua. Adjacent vacua are separated by a canonical distance of at least $\xi_1$, where $\xi_1^2$ is the smallest eigenvalue of the corresponding kinetic matrix. The difference in energy density between the two vacua is constrained by the scale of the subleading potential, while a typical field trajectory interpolates through a potential of scale $\Lambda_{\mathscr Q}^4$. By demanding $B\gtrsim 10^3$ we then find a rough bound for the scale of the smallest axion decay constant,
\be\label{tunnelingexpp}
\xi_1 \gtrsim {\Lambda_{\text{r}}^3\over  \Lambda_{\mathcal Q}^2}\,.
\ee
Gravitational contributions to the decay rate are negligible in the relevant regime. Therefore, theories that satisfy the constraint in (\ref{tunnelingexpp}) are expected to have vacua that are stable on time scales long compared to the age of the universe.

Following Coleman-De Luccia decay to a vacuum with sufficiently small cosmological constant, the universe exhibits negative spatial curvature and may undergo slow-roll inflation. Since the vacua within one axionic band are spread over a range of $\Lambda_{\text{r}}^4$, the tunneling process to our current vacuum will leave the axions at an energy of about $\Lambda_{\text{r}}^4$, evading the empty universe problem for sufficiently large energy bands. To understand this important point, consider tunneling from a penultimate vacuum at $\phi^*_{\alpha,\mathbf n^\prime}$ to our current vacuum with very small cosmological constant $V(\phi^*_{\alpha,\mathbf n})\ll \Lambda_r^4$. For simplicity, let us consider only one subleading term in the axion potential with dynamical scale $\Lambda_r^4$. The degeneracy between the two vacua is lifted by the subleading potential, 
\be
\delta V_{{\alpha};\,\mathbf n,\,\mathbf n^\prime}\approx\Lambda_r^4\cos\left(2\pi Q_\text{\text{r}}\boldsymbol B_{\mathscr Q}\mathbf n^\prime+\tilde{\delta}_\alpha\right)\,,
\ee
where we neglected the small vacuum energy in the final vacuum and $\tilde{\delta}_{\alpha}$ is a phase. The sum inside the cosine consists of $N$ order one terms. Therefore, neighboring vacua, will differ by about $\Lambda_r^4$ in energy density. Finding a vacuum with energy density in a given range is a difficult problem: we need to test an exponential number of vacua before finding one with a suitably small  energy. Since tunneling between widely separated vacua is exponentially suppressed, the energy released during the final transition is of order $\Lambda_r^4$. This feature is due to the multi-dimensionality and absent in theories with a single axion. This energy density can lead to a subsequent phase of slow-roll inflation, while the tunneling event may give rise  to  observable features \cite{Kleban:2012ph,Bousso:2014jca}. 

It is curious to point out that when considering the Lagrangian (\ref{Lone}) with $N=500$ axions and $P=N$ leading contributions of scale $\Lambda_a\sim 0.1\, \M$, some subleading contributions at the GUT scale and $2 \%$ non-vanishing order one entries in the charge matrix, we expect a sufficient number of vacua to account for the observed smallness of the cosmological constant and, at the same time, large field axion inflation via kinetic alignment, which would solve the flatness problem \cite{Bachlechner:2014hsa,Bachlechner:2014gfa}. However,  the inflationary dynamics are highly sensitive to heavy fields, so it is not clear if this model can be embedded in a consistent theory of quantum gravity.

\section{The strong CP Problem}
The action of QCD famously contains a CP-violating term that is proportional to the Yang-Mills instanton number,
\be
\delta S={\theta_{QCD}\over 8\pi} \int d^4 x~\text{Tr} (F_{\mu\nu}\tilde{F}^{\mu \nu} )\,.
\ee
By measuring the electric dipole moment of the neutron one obtains an upper bound on the coupling parameter $|\theta_{QCD}|\lesssim 10^{-10}$. The smallness of this dimensionless parameter constitutes the strong CP problem. One of the most compelling approaches to this problem is to promote the coupling constant to a dynamical field with a continuous shift symmetry that is broken to a discrete shift symmetry via its coupling to the QCD anomaly \cite{Peccei:1977hh}. This generates a potential of the form
\be
V_{\text{QCD}}= \Lambda_{QCD}^4\cos(2\pi \theta_{QCD})\,,
\ee
where $\Lambda^4_{QCD}\approx f_{\pi}^2m_\pi^2$ . In the absence of  additional couplings the axion is dynamically stabilized at $\theta_{QCD}=0$. However, upon embedding QCD in a theory of quantum gravity, such as string theory, there are additional contributions to the QCD axion potential. These high energy contributions  take the form $\delta V\sim \Lambda^4_{UV}\cos(2\pi \theta_{QCD}+\delta)$, where $\delta$ is some order one phase, set at a high energy. Since the expectation value for the QCD axion is now dominated by the high-energy effects this leads to a large CP violating phase $\delta$ \cite{Banks:1996ea,Svrcek:2006yi}.
This is precisely the situation we have when considering a general coupling between the axions $\theta^i$ in Lagrangian (\ref{Lone}) to QCD, with charges $q^{QCD}$. However, as argued above, in a multi-axion theory, we  expect a large number of possible values for the CP violating phase. While this theory does not dynamically favor CP conservation, the band structure is  sufficient to accommodate a small CP violating phase.

\section{Conclusions}
We studied the vacuum distribution of a theory containing multiple axions. In general, it is difficult to precisely determine the number and location of vacua in a multi-dimensional potential. In order to simplify the problem, we separated the leading and the sub-leading contributions to the axion potential: the leading terms determine the vacuum periodicity, while sub-dominant contributions break this discrete shift symmetry and lift the vacuum  degeneracy. In the universal regime, the number of discrete vacua scales exponentially with the number of axions. The splitting of the degenerate energy levels is small compared to the typical scale of the potential energy, giving rise to energy bands containing stable vacua. If one of the energy bands spans zero energy, there exist  vacua with very small cosmological constant. In an eternally inflating universe, this vacuum is populated which leads to a small observed vacuum energy.

The simple observations made in this work may have profound implications. 
We demonstrated that a sufficiently complex landscape to accommodate the cosmological constant and a small CP violating phase can be realized in a large class of four dimensional effective theories. The energy density of domain walls that arise during vacuum transitions can be low enough to decouple from unknown, heavy degrees of freedom. The only requirement for a large number of vacua is that the leading instanton charges do not form a primitive basis for the integer lattice. An intriguing observation is that the simplest multi-axion model that  accommodates the smallness of the cosmological constant  also gives rise to an extended period of large-field inflation via kinetic alignment. Finally, our work motivates a more detailed study of the axion sector in flux compactifications and provides a new framework to pursue de Sitter vacua in string theory.

\section{Acknowledgements}
I  would like to thank Raphael Bousso, Frederik Denef,  Michael Douglas, Cody Long, Liam McAllister, Eve Vavagiakis, Erick Weinberg, and Claire Zukowski for useful discussions.
This work was supported by DOE under grant no. DE-SC0011941.

\bibliography{References}

\end{document}